\documentclass[10pt]{article}
\usepackage{amssymb,bbm,amsmath,stmaryrd,subfig,color,mathrsfs,graphicx,pxfonts}
\usepackage[paperwidth=16cm,paperheight=24cm,textwidth=10cm,textheight=10cm,top=2cm,left=2cm,right=2cm,bottom=2cm]{geometry}
\usepackage{hyperref}
\definecolor{refcolor}{rgb}{0.3,0.3,0.7}
\hypersetup{
    pdftitle={The role of phase interface energy in martensitic transformations: a lattice Monte-Carlo simulation},
    pdfauthor={Vladislav A. Yastrebov},
    pdfsubject={Martensitic transformations},
    pdfcreator={LaTeX, Kile},
    pdfkeywords={martensitic transformation, interface energy, statistical mechanics, lattice Monte-Carlo simulation},
    pdfnewwindow=true,
    colorlinks=true,
    linkcolor=refcolor,          
    citecolor=refcolor,          
    filecolor=refcolor,      
    urlcolor=refcolor           
}

\usepackage{setspace}

\newcommand{\stress}{\boldsymbol\sigma}
\newcommand{\strain}{\boldsymbol\varepsilon}

\newcommand{\hath}{\,\hat{\!\mathcal{H}}}
\newcommand{\vecmu}{\mathbf{\boldsymbol{\mu}}}

\title{The role of phase interface energy in martensitic transformations:\\a lattice Monte-Carlo simulation}
\author{Vladislav A. Yastrebov$^a$ \qquad Michael Fischlschweiger$^{a,b}$\\
Georges Cailletaud$^a$ \qquad Thomas Antretter$^c$}

 \date{\footnotesize$^a${\it Centre des Mat\'eriaux, MINES ParisTech, CNRS UMR 7633, BP 87, F 91003 Evry, France}\\
                    $^b${\it Materials Center Leoben Forschung GmbH, Roseggerstrasse 12, 8700 Leoben, Austria}\\
		    $^c${\it Institute of Mechanics, University of Leoben, Franz-Josef-Strasse 18, 8700 Leoben, Austria}
}

\begin{document}
 
\maketitle

\begin{abstract}

To study martensitic phase transformation we use a micromechanical model based on statistical mechanics.
Employing lattice Monte-Carlo simulations and realistic material properties for shape-memory alloys (SMA), 
we investigate the combined influence of the external stress, temperature, and interface energy between the austenitic and martensitic phase on the 
transformation kinetics and the effective material compliance. 
The one-dimensional model predicts well many features of the martensitic transformation that are observed experimentally.
Particularly, we study the influence of the interface energy on the transformation width and the effective compliance.
In perspective, the obtained results might be helpful for the design of new SMAs for more sensitive smart structures and more efficient damping systems.

%
\end{abstract}





\section{Introduction}

Displacive solid-to-solid phase transformations (also termed as martensitic phase transformations) are accompanied by remarkable changes in the material behavior making these materials important for the design of intelligent structures \cite{Bhattacharya2004, Bhattacharya2003, Aaltio2008, Otsuka2011}.
It has been highlighted that these changes are strongly dependent on the austenite-martensite (A-M) interface energy 
which in the end controls the material's applicability for sensors and damping elements in structures~\cite{Waitz2007,Humbeeck2003,Basu1999}.
Therefore, the role of the interface energy in the transformation kinetics and consequently in the material behavior 
has drawn an increasing interest in condensed matter physics \cite{Bertoldi2007,Waitz2009,Lei2010, Artemev2000}.
In the past, mainly continuum based models have been developed \cite{Levitas2002,Petryk2010}.
Recently, a phase field approach has been applied to characterize the effect of geometric incompatibilities 
and crystal symmetries (twins and habit plane formation) on the A-M interface energy \cite{Lei2010}.
A consistent framework to incorporate surface tension
and energy in the Ginzburg-Landau theory for multivariant martensitic phase
transformation has been developed by~\cite{levitas2010prl,levitas2011ijmr} for modeling coherent interfaces.
Lately, an approach based on statistical mechanics of lattice systems linking the microscopic state and the macroscopic behavior 
has been applied to describe consequences of martensitic transformations on the mechanical  behavior of materials~\cite{Oberaigner2011}.
In this approach, the phase interface energy can easily be taken into consideration.
The interface energy contains chemical and mechanical contributions, where the latter are due to elastic energy stored in the vicinity of the interface between the austenitic and martensitic phases \cite{Waitz2007}. 
So the interface energy considered here does not present a microscopical (local) quantity, 
but should be considered at a mesoscopic scale as an integral measure of the potential interface between austenite and martensite phase.
This interpretation implies higher values of the interface energy than is usually used in micromechanical models.  
A straight atomically perfect interface has a minimum energy. The less perfect the interface, the higher both the stored energy and the transformation barrier \cite{Lei2010}.

\begin{figure}[ht!]
\includegraphics[width=1\textwidth]{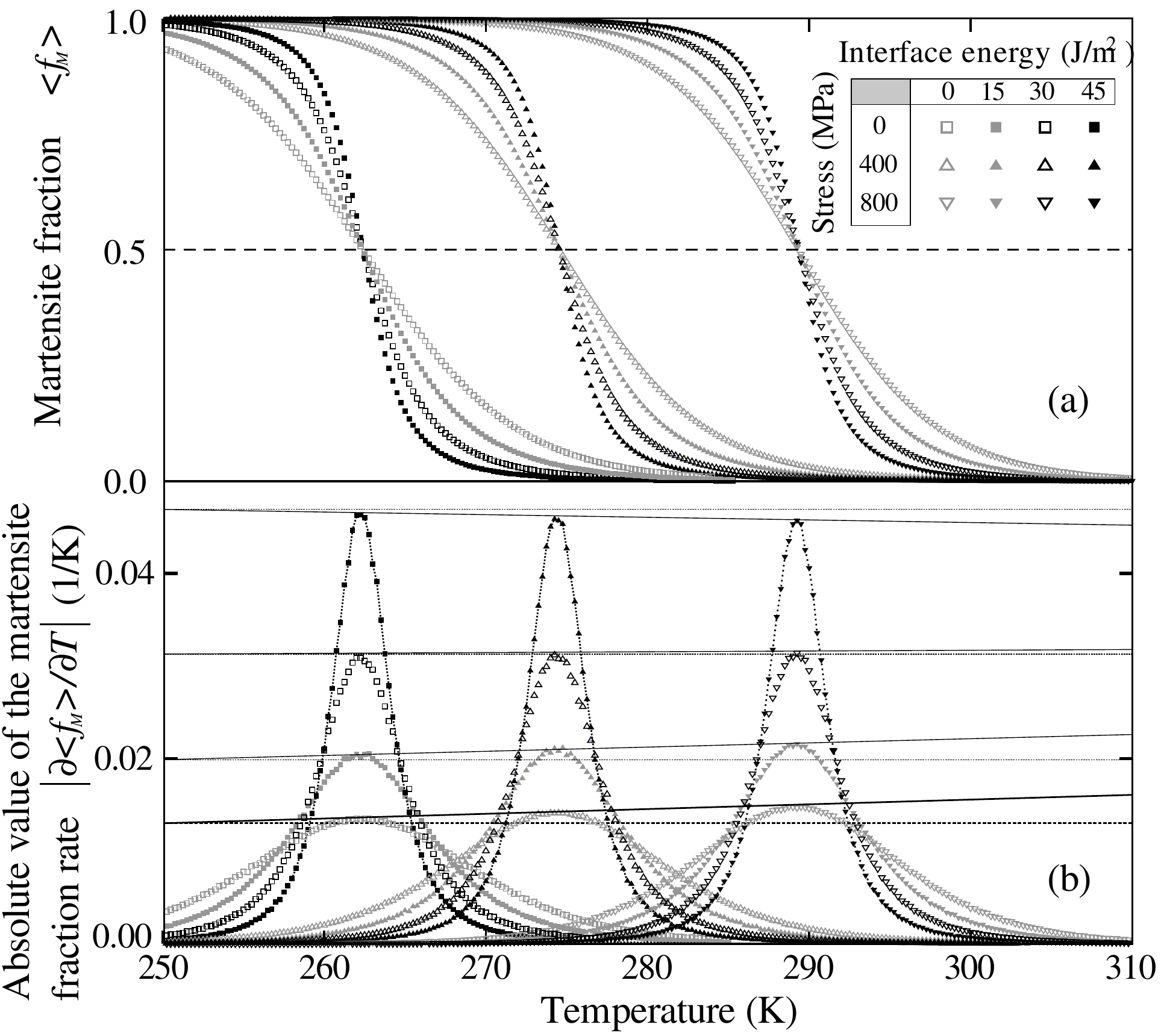}
\caption{\label{fig:1} The temperature-induced transformation kinetics depends strongly on the external applied stress and the austenite-martensite interface energy. 
(a) -- The martensite phase fraction $\langle f_{_M} \rangle$ evolution with temperature for different interface energies and external stresses and 
(b) -- the absolute value of the martensite fraction rate $|\partial \langle f_{_M} \rangle / \partial T|$ are depicted. 
The transformation zone narrowing for high interface energies and the increase (decrease) of the peak rate value are observed.}
\end{figure}

In the current article we employ a lattice Monte-Carlo simulation  
for martensitic phase transformations in analogy to the techniques used in magnetic systems~\cite{Landau2009}.
But in contrast to the latter, we use a specific Hamiltonian proposed in~\cite{Oberaigner2011}.
We discuss the influence of the interface energy on the width of the transformation zone and material's compliance depending both on stress and temperature paths.


\section{Theory and statistical model}

In the model we assume that the center of each single crystal (microscale) occupies a site of an $L$-component one-dimensional lattice (macroscale).
The microstate specifying variable is a vector $\boldsymbol{\mu}$, each component $\mu_i, i \in [1,L]$ can be either A or M. 
The space of microstates for the prescribed stress-temperature state is labeled by $\wedge$. 
We use the constant pressure (isobaric) distribution, the pressure is interchanged by the Cauchy stress tensor $\stress$ \cite{Lavis2010}. 
The partition function $Z$ and the Hamiltonian $\hath$ are defined as
\begin{equation}
 \begin{split}
   &Z = \sum_{\wedge} \exp\left[ - \beta \hath (T,\vecmu)\right],\\
&\hath (T,\vecmu) = E(T,\vecmu) - \stress : \,\tilde{\!\strain}( \vecmu ),
 \end{split}
 \label{eq:1}
\end{equation}
where $\beta$ is a Lagrange multiplier $\beta=1/(kT)$, where $k$ is the modified Boltzmann constant which determines the relative size of the cells, and $T$ is the temperature; the total internal energy
\begin{equation}
E(T, \vecmu) = \sum_{i}^{L} U_{\mu_i}(T) + 1/2\!\!\!\!\sum_{i=1,|i-j|=1}^{L} U_{\mu_i,\mu_j}^I.
\label{eq:2}
\end{equation} 
is a combination of the internal $U_{\mu_i}(T)$ and
the interface $U_{\mu_i,\mu_j}^I$ energy densities between neighbor sites.
In a first assumption, the interface energy is a temperature independent quantity.
It is worth noting that the internal energy densities of the martensite and austenite have only
one intersection point on the considered temperature interval $T\in[T_0; T_1]$, moreover,
for $T=T_0$, $U_{_{A}} > U_{_{M}}$ and for $T=T_1$ $U_{_{A}} < U_{_{M}}$, in our case $T_0 \approx 2$ K, $T_1 \to \infty$.
In a single martensitic variant system the interface energy between sites is given by 
\begin{equation}
U_{\mu_i,\mu_j}^I = U_{_{AM}} \left( 1-\delta_{\mu_i}^{\mu_j}\right),
\label{eq:2a}
\end{equation}
where $U_{_{AM}}$ is the austenite-martensite interface energy per unit surface and $\delta_{\mu_i}^{\mu_j}$ is the Kronecker delta. 
The term $\stress : \tilde\strain(\vecmu)$ in Eq.~(\ref{eq:1}) takes into account the energy through an external mechanical stress $\stress$ and 
the strain density tensor $\;\tilde{\!\strain}( \vecmu )$ given by
\begin{equation}
  \tilde{\!\strain}( \mu_{_A} ) = \mathbf{S}_{_A}\!:\stress / \eta_{\mu_A},\quad \tilde{\!\strain}( \mu_{_M} ) = \left(\mathbf{S}_{_M}\!:\stress + \strain^{\mbox{\tiny tr}}\right) / \eta_{\mu_M}
\label{eq:3}   
\end{equation}
for austenite $\mu_{_A}$ and martensite $\mu_{_M}$ sites, respectively, where $\strain^{\mbox{\tiny tr}}$ is the transformation strain tensor for the given variant of the martensite, $\mathbf{S}$ is the fourth-order compliance tensor from the theory of linear elasticity and $\eta_{\mu_i}$ is the molar density of a pure state.
The decomposition of the total strain is made according to the framework of small strain theory.
In a first assumption, the model omits thermal strains and stress fluctuations.
Further information about the effective Hamiltonian formulation can be found in~\cite{Oberaigner2011}.

Here, we employ a Monte-Carlo simulation using the Metropolis algorithm~\cite{metropolis1953} for a one-dimensional A-M lattice system with fully periodic boundary conditions.
The lattice site interfaces are simulated as soft fluctuating objects within the lattice.

\section{Description of computation}

One of the properties of interest in martensitic phase transformation is the evolution of the martensite fraction $\langle f_{_M}\rangle$.
In a Monte-Carlo simulation, this quantity can be computed as follows: by reaching the equilibrium at time $t_0$ for a given stress-temperature state, one obtains
the martensite fraction as an average of the instantaneous martensite fractions $f_{_M}(t^{\mbox{\tiny MC}},T,\stress)$ over interval $\Delta t$ of the discrete and dimensionless Monte-Carlo time $t^{\mbox{\tiny MC}}$:
$$
  \langle f_{_M}(T,\stress)\rangle = \frac{1}{\Delta t} \sum\limits_{t^{\mbox{\tiny MC}}=t_0}^{t_0+\Delta t} f_{_M}(t^{\mbox{\tiny MC}},T,\stress)
$$
We found that the results converge with an increasing number of lattice sites $L$, i.e.
$$\forall\varepsilon > 0, \exists\,L: \forall l > L: \left\|\langle f_{_M}(T,\sigma)\rangle_{{l}} - \langle f_{_M}(T,\sigma)\rangle_{{L}}\right\| \le \varepsilon.$$
For $\varepsilon=1\%$ it is  found that $L\approx7$.
The symbol $\| \dots  \|$ denotes here the maximum norm $L^\infty$.

\begin{figure}[ht!]
\includegraphics[width=1\textwidth]{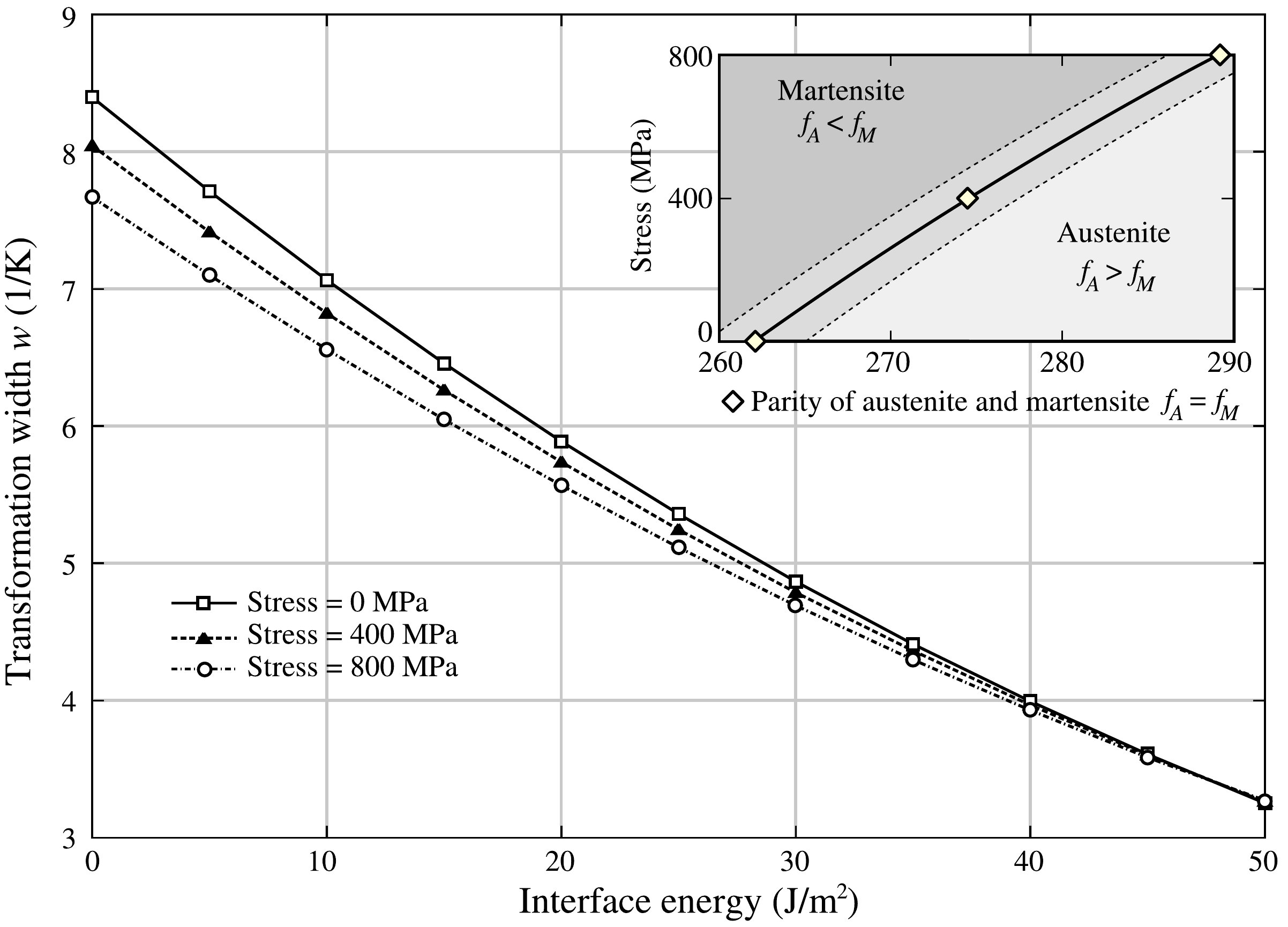}
\caption{\label{fig:2} A smooth decrease of the broadness parameter $w$ with increasing interface energies; the influence of the 
external stress is relatively small; the shift of the parity point (mean transformation temperature, for which $f_{_A} = f_{_M}$) under applied stress is depicted in the inset.}
\end{figure}

The lattice system under consideration does not exhibit a critical slowing down, thus  
the simple Metropolis algorithm is applicable for a proper prediction of the phase transition, 
in contrast to magnetic systems which require cluster algorithms to avoid this slowing down phenomenon (e.g., Swendson-Wang algorithm \cite{Landau2009,Swendsen1987}).
The input data for our model include molar densities $\eta_{_A},\eta_{_M}$, 
elastic properties $\mathbf S_{_A},\mathbf S_{_M}$, 
evolution of the internal energy densities $U_{A}(T), U_{M}(T)$ of the pure phases (which can be found from specific heat capacities $c_{_A}(T),c_{_M}(T)$~\cite{Oberaigner2011}),
interface energy densities between austenite and martensite phases $U^I_{_{AM}}$ as well as between different martensitic variants (if they are present)
and the crystallographic transformation strain $\strain^{\mbox{\tiny tr}}$ of the martensitic variants. 
Herein, typical material data for a shape memory alloy are taken from~\cite{Oberaigner2011}:
Young's moduli $E_{_A} = 56$ GPa, $E_{_M}=26$ GPa, Poisson's ratios $\nu_{_A} = \nu_{_M} = 0.3$, molar densities $\eta_{_A} = \eta_{_M} = 9.7\times 10^4$ mol/m$^3$,
transformation strain $\varepsilon_{11}^{\mbox{\tiny tr}} = 2\%$, internal energy densities 
$U_{A}(T) = 10 + 7.66 T + 0.033 T^2$ J/m$^3$,
$U_{M}(T) = 20 - 0.25 T + 0.067 T^2$ J/m$^3$ and the interface energy density $U_{_{AM}}$ varies up to $50$ J/m$^2$, whereas the 
Lagrange multiplier $\beta = 7.24/T$.
For the lattice simulation we prescribe the number of sites $L=10$ and 
the Monte-Carlo time period $\Delta t$ over which the data are averaged when the simulation reaches a quasi-equilibrium state, in our case
$\Delta t^{\mbox{\tiny MC}}=10^8$ per each stress-temperature state. In all simulations the temperature step is $\Delta T = 0.1$ K.

The simulation is started from a pure austenite state at $T=310$ K for a given external stress $\stress$.
According to the Metropolis algorithm a random site is flipped either from austenite to martensite or vice versa.
If a random number $r$ uniformly distributed in $[0;1]$ is smaller than the probability resulting from the change of the Hamiltonian 
$r < \exp\left(-\beta\Delta\hath (T)\right)$, then the change of the state is accepted, otherwise not.
The change of the Hamiltonian is given by $\Delta\hath (T) = \hath (T,\vecmu^1)-\hath (T,\vecmu^0)$, 
where $\vecmu^1$ and $\vecmu^0$ are configurations after and before the toggling, respectively.
The procedure is repeated $10^8$ times, and the average values over time are computed.
Then the temperature is decreased by $0.1$ K and the procedure is repeated again. Moreover, the change of the average martensite fraction is computed with
respect to the change of the temperature (see Fig.~\ref{fig:1}).
The increments are repeated up to $T=190$ K.

\section{Results}

\subsection{Width of the transformation zone}

The simulation of thermally induced transformations shows a considerable influence of the interface energy on the transformation kinetics. 
This result holds for all stress states in the elastic regime.
The higher the interface energy the faster the transformation proceeds, see Fig.~\ref{fig:1}.
This narrowing of the transformation zone implies an increased phase stability.
However, the temperature of parity of phases $f_{_A}(T^\ast) = f_{_M}(T^\ast)$ is a function of the internal energies and the external stress 
and does not depend on the interface energy.
To characterize the width of the transformation zone we introduce a parameter $w\;[1/K]$, defined as a mean squared
derivative of the martensitic fraction with respect to the temperature
\begin{equation}
  w = \int\limits_0^\infty \left[\frac{\partial \langle f_{_M} \rangle}{\partial T} - \left\langle\frac{\partial \langle f_{_M} \rangle}{\partial T}\right\rangle\right]^2\,dT,
  \label{eq:4}
\end{equation}
where $\left\langle \partial \langle f_{_M} \rangle /\partial T \right\rangle$ is the mean slope of the martensitic phase fraction.
Fig.~\ref{fig:2} illustrates a smooth decrease of the width parameter $w$ with the interface energy. 
Note that the width is higher for smaller stresses if the interface energy is below $45-50$ J/m$^2$,
at this value the curves for different stresses collapse.

\subsection{Effective compliance}

\begin{figure}[ht]
\includegraphics[width=1\textwidth]{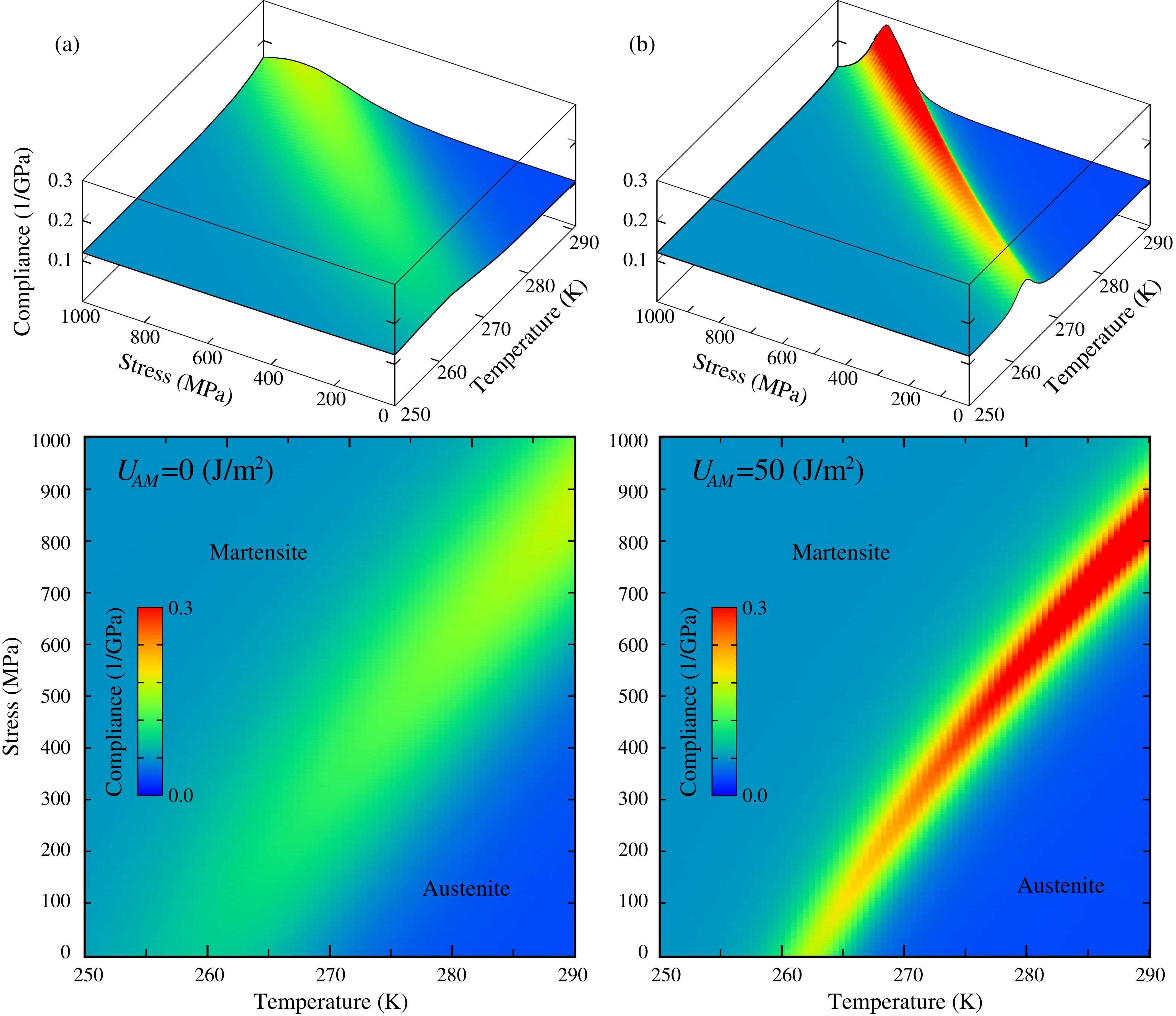}
\caption{\label{fig:3} Compliance contour-plot for a material undergoing a martensitic phase transformation in the stress-temperature space (each point $10^8$ Monte-Carlo time steps); an increase in the interface energy results in an increase of the compliance accompanied by a transformation zone narrowing (a) -- $U_{_{AM}}=0$, 
(b) -- $U_{_{AM}}=50 J/m^2$.}
\end{figure}

To demonstrate how the interface energy affects the damping properties of materials obeying martensitic transformations, we remark that
the deformation wave speed $v$ (in a one-dimensional case only longitudinal waves are present) is inversely proportional to the square root of the product of the density and the compliance
$$
1/v \sim \sqrt{\rho C(U_{_{AM}})},
$$ 
where $C = \partial \varepsilon/\partial \sigma$. 
For shape memory alloys the density is almost independent from the phase, whereas the compliance increases significantly within the transformation region 
as shown in Fig.~\ref{fig:3}.
For a given temperature the peak compliance increases with increasing interface energy.
Thus, a shock wave, which is  capable of initiating a stress-induced transformation, slows down within the transformation zone. 
The higher the interface energy, the more pronounced is this deceleration process.
A further important factor for damping properties is the hysteretic behavior of the material, therefore it is the  future step in our study to introduce a physically meaningful hysteresis mechanism. 

\section{Conclusions and perspectives}

Using one-dimensional Monte-Carlo lattice simulations based on a statistical micromechanical model, 
we demonstrated for different external stresses a smooth increase in the martensite phase fraction with decreasing temperature. 
The width of the martensite transformation is shown to depend on the austenite-martensite interface energy: the higher this energy, the sharper the transformation.
Note that to demonstrate better this trend and to partly compensate the one-dimensionality of our system, we deliberately exaggerated the interface energy.
By analyzing the transformation in stress-temperature space, we show the influence of the interface energy on the effective compliance 
of the material subjected to a martensitic transformation. 
In agreement with the stress-temperature curve generally observed experimentally for the onset of the martensite transformation~\cite{Lagoudas2008b},
we obtain an almost linear relation between stress and temperature for the parity curve (inset in Fig.~\ref{fig:2}).

Our results suggest a possible direction in the improvement of shape-memory alloys aimed at increasing the interface energy between the martensite and the austenite phases.
Use of materials with a sharper martensite transformation (owing to a high interface energy) 
will result in more sensitive and energy-efficient smart systems: switches, actuators, etc. 

For an investigation of the entire properties and quantitative comparisons with experiments it is inevitable to simulate 
at least a two-dimensional lattice system including multiple martensitic variants and internal stresses.
Interface energies between martensitic variants [as in~\cite{levitas2011ijmr}] and the choice of the variant under complex external loads and temperatures are also under consideration.
Although it is easy to extend the Monte-Carlo simulations for an arbitrary dimension of the lattice, the incorporation of internal stresses in the formulation 
is not trivial and the work in this direction is in progress. Moreover, we are working on a thermodynamically coherent introducing of the latent heat 
and the internal friction \cite{Fukuhara2002prb,Waitz2007} in the system in order to reproduce a transformation hysteresis.

\section{Acknowledgment}

Financial support by the Austrian Federal Government within the research activities of the K2 Competence Center on Integrated Research in Materials, Processing and Product Engineering is gratefully acknowledged.


\end{document}